# Quantum correlation tests at cosmic distances

Thomas Durt[1], Jean Schneider[2]
1: Aix Marseille Univ, CNRS, Centrale Marseille, Institut Fresnel
13013 Marseille, France
2: Paris Observatory - LUX, 5 place Jules Janssen, Meudon 92190, France

**ABSTRACT.**
*It is commonly accepted that the results of measurements simultaneously realized over two entangled subsystems are statistically correlated instantaneously regardless of the distance between them. In accordance with Bell's theorem, everything happens in such measurements as if there was a correlation propagating at infinite speed between the two subsystems. These correlations have been so far verified experimentally up to a distance of 1200 km. We discuss the interest and feasibility of extending this distance to 390,000 km, thus gaining a factor of 300. The idea is to install one of the polarimeters on the Moon, with the other on Earth. Such an experiment would provide a new test of Quantum Physics and allow to put higher constraints on alternative theories and interpretations. We also discuss the possibility to violate Bell's inequalities beyond Earth-Moon distance.*

## 1 Introduction.

In Louis de Broglie's idea of associating waves with particles of matter, these waves were physical waves. What he did not foresee was that the result of an observation is random. For example, the wave associated with a photon is extended, whereas it is only detected at one point, which cannot be predicted in advance. At the same time, in the minds of the early creators of Quantum Mechanics (or Matrix Mechanics), the only tangible reality consists of observables, with the mathematical symbols associated with waves being merely mathematical intermediaries intended to predict the results of measurements and their probability in a given experimental context. While pragmaticallyagreeing with this point of view, given the undisputed experimental successes of the majority view of quantum mechanics, Louis de Broglie was never truly satisfied with it, proposing, for example, the theory of the "double solution" [1, 2, 3, 4, 5].

One of the problems with the orthodox interpretation is that the result of any measurement is random, obeying Born's rule. This led Von Neumann to introduce the ad hoc postulate of the "(probabilistic) reduction of the wave packet", which cannot be described by a Schrödinger equation because of the non-deterministic, irreversible in time, and nonlinear nature of the collapse process. Consequently, an observation process cannot be described as a physical process within the framework of the theory, i.e. as an interaction between the measured object and a measuring device, which strikes the mind of a physicist.

This situation was uncomfortable for Louis de Broglie who, under the impetus of physicists such as David Bohm, Georges Lochak and Jean-Pierre Vigier, considered alternative interpretations of wave mechanics [2].
One of them is the so-called the de Broglie-Bohm interpretation, in which particles follow deterministic and continuous trajectories at all times. As already remarked by de Broglie in 1927 [1] however, the dynamics must be formulated in the configuration space in the case of entangled subsystems, which opens the door to non-locality, a feature that de Broglie disliked during all his life. He actually tried to reformulate the de Broglie-Bohm dynamics in real, 3 D, physical space with Andrade e Silva [6, 7, 8], but this approach is nowadays nearly entirely forgotten. He also considered seriously the possibility of the existence of a subquantum ether (denoted by him the "quantum thermostat" [9]), an idea also developed by Vigier [10, 11]. What is interesting with these realistic approaches is that they suggest that quantum correlations propagate in the real, 3D, space, and not in the configuration space.

From this point of view, the speed of quantum correlations is not necessarily infinite, as it would be in the configuration space. In the present paper we discuss the experimental lower bounds on the speed of quantum correlations achieved so far and propose to win a factor 300 over the distance over which quantum correlations have been tested.

Another interest of our proposal is to test whether quantum correlations themselves would disappear at large distances, an idea also shared by Louis de Broglie [8, 12] before Aspect's experiment [13] was realized. It is indeed a very intriguing property of entangled systems that their correlations are in principle independent on their distance, which is a deeply non-classical feature if we think of classical interactions such as gravity and electro-magnetism. From this point of view any gain in the distance over which we test the presence of non-local correlations is a plus. It is actually in the context of QKD that maximal distances of transmission of entangled photons have been obtained so far [14]. Last but not least, exchanging entangled photons between Earth and Moon also paves the way to quantum key distribution (QKD) at planetar scales, which is a stimulating technological challenge.

## 2 Quantum non-local correlations

One consequence of the projection postulate, paradoxical in terms of everyday intuition, is the remote (statistical) correlation between measurement results. The problem originated with the "EPR paradox" of Einstein, Podolski and Rosen, who produced an argument intended to experimentally disprove pure probabilism [15]. This led John Bell to show that, in certain situations (figure 1), the hypothesis that their values are determined before they are measured contradicts the axioms of quantum mechanics [16]. Realism, in the spirit of Bell's original article and EPR's paper as well, means that individual particles are entities that possess their own properties, which they carry with them. More precisely, the statistical correlation of the measurement results of the two complementary quantities, namely the projections J1a and J2b of the angular momentum on any axes a and b of two particles 1 and 2 cannot be the same if we assume that their value pre-exists before measurement as predicted by orthodox quantum theory. Mathematically, this translates into Bell's inequalities for the polarisation of a pair of photons emitted from a spin-0 atom.

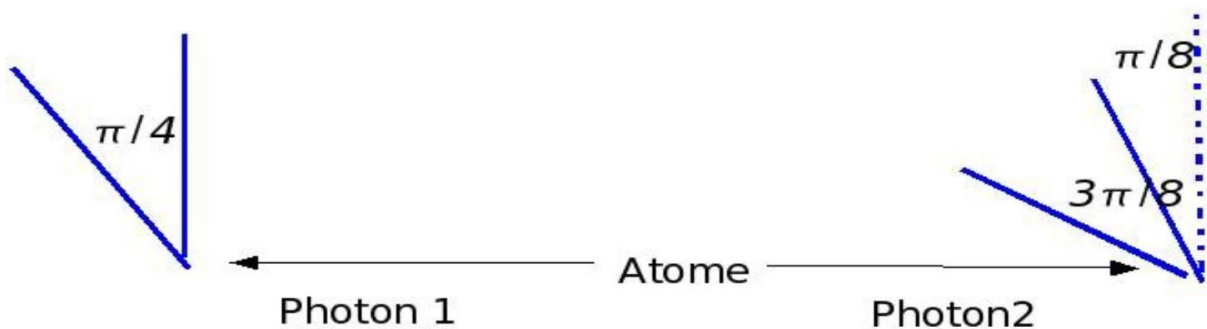

Figure 1: Statistically correlated projective measurements performed on polarisation states of two entangled photons.

In the "realistic" hypothesis where their value pre-exists before measurement,

$$C(0, \pi/8) - C(0.3\pi/8) + C(\pi/4, \pi/8) - C(\pi/4, 3\pi/8) \leq 2,$$

where C(a,b) is the statistical correlation factor of the measurement results of J 1,a and of J 2 ,b . On the contrary, quantum mechanics predicts

$$C(0, \pi/8) - C(0.3\pi/8) + C(\pi/4, \pi/8) - C(\pi/4, 3\pi/8) = 2\sqrt{2}$$

which violates the inequality above. Several experiments confirmed the violation of such Bell's inequalities up to a distance of 1200 km [14]. It reveals the existence of "spooky action at a distance" in Nature.

## 3 Estimating a lower bound on the "speed of collapse"
### 3.1 Discussion.
If the violation of Bell's inequalities nullifies the possibility of a scenario à la EPR where the measurement outcomes are "written in advance", it does not rule out however the possibility of a mechanisms through which the subsystems communicate faster than light [17]. de Broglie-Boh dynamics provides for instance an example of such a communication, mediated by the phase of the entangled wave function of the system. In this case, correlations propagate at infinite speed relatively to a quantum ether which plays the role of a privileged frame. It is common to choose this frame in such a way that it is at rest relatively to the lab. but other choices are possible; for instance one could choose a frame comoving with the center of Earth, or with the Sun. Whichever privileged frame we choose, it is impossible to discriminate between de Broglie-Bohm's interpretation and the orthodox theory. One could imagine however that the subsystems communicate at finite speed. If things happen in this way, standard quantum predictions could possibly get violated during experiments realized at cosmic distances as we will discuss soon.Testing quantum correlations.

### 3.2 Lower bound on the speed of quantum non-local correlations.
An intuive way to estimate the speed of quantum correlations is to divide the distance between the measurements performed on two entangled subsystems A and B by the absolute value of the time interval separating them.

However, when the measurements are simultaneous, this speed goes to infinity and no lower bound can be inferred from the experiment.

It is not realistic however to assume that a measurement is an instantaneous process. Each measurement requires a certain non zero time to be achieved. It is thus possible to explain correlations if the two subsystems "communicate" during the time required to perform the measurement. The reasoning goes as follows: if the time required to perform simultaneous measurements is short enough and the distance between the subsystems is large enough, and if moreover the speed of communication between them is too slow, their communication will begin after the measurements are performed, which implies that the correlations will no longer violate Bell's inequalities. It is for instance in this way that some years ago Nicolas Gisin and his team derived a lower bound of the order of $7 \times 10^6$ c on the speed of quantum correlation[1].

---

[1] In ref. [18], on can read the following: ...*In 1999 we performed an experiment in which careful fiber length and chromatic dispersion adjustments provided a timing accuracy of ± 5 ps over a distance of 10.6 km, setting the lower bound: vQI ≥ 32 $10^7$ c, where c denotes the speed of light....*

To do so he divided the distance between the two places where the polarisations of the entangled photons were simultaneously measured by the duration of such a measurement. In Gisin's experiment, this distance was equal to 10 kilometer while the duration was estimated to be 5 picosecond[2], which provides a lower bound of the order of 700000 times c.

In what follows we shall assume that, whichever experimental configuration we consider, the duration of a measurement is the same as in Gisin's experiment. Other choices are possible of course. For instance one could as well replace the duration of the measurement by the time elapsing between the setting of the measurement basis and the end of the measurement. It can also occur that the electronics is not as fast as in Gisin's experiment (2,5 picosec) or that the uncertainty on the timing of the subsystems is not as short as 2,5 picosec. For reasons of simplicity we shall not consider all these possibilities here. Of course, we are free to recalibrate the value of the duration of the measurement chosen here (5 picosec) taking account of the real conditions in which the experiment is realized but this does not fundamentally change our way to tackle the problem.

In our estimate of the lower bound on the speed of quantum correlation we shall however slightly modify the choice of Gisin. Indeed, instead of the distance "in straight line" between the subsystems we will rather consider the double of the maximal length of the trajectory followed by a subsystem since it was created by the source[3], where we maximize over the subsystems. The reason for this choice is that in an approach where one considers that correlations behave as waves travelling through an hypothetical quantum aether [17, 20], it is more natural to assume that these waves follow the trace left by the entangled particles in this medium. This idea is not new; for instance Buonomano [22, 23] proposed that quantum particles impregnate space along their trajectories in the same way that ants communicate with each otehr through chemical signals (pheromons).

Finally, we shall keep the same estimate of the speed of quantum correlations even when the two measurements are not simultaneous. The reason therefore is that we assume, which is a fair and natural assumption, that when a subsystem undergoes a measurement, and that a "collapse wave" departs from it and follows the path joining other entangled subsystems, this wave does not possess any information about where and when measurements are performed on the entangled partner(s).
The speed of communication must thus be maximal in order to comply with the most constraining scenario, the one in which simultaneous measurements are performed. All our assumptions listed above lead to an operational way to estimate (a lower bound on) the speed of quantum correlations : this speed is equal to the double of the maximum of the distance between the source and one of the detectors divided by 5 picosecond as explained above. Obviously, as explained below, increasing the distance over which entangled photons exhibit non-local correlations imposes stronger bounds on the minimum of the speed of quantum correlations, which is the main motivation of the present proposal.

______________________________

[2] This duration was estimated by summing the uncertainty on the timing of the photons, with the time required for a "click" to occur (this is the response time of the single photon detectors used in the experiment. Both times were estimated by Gisin to be of the order of 2,5 picosecond.

[3] This definition is operational in the context of the present paper because we only consider experiments where entangled particles are issued from a common source. In the case of entanglement swapping [19], for instance when A is initially entangled with B, and C with D, and that a Bell measurement is performed on the BC pair in order to entangle A with D, one can easily convince one self that we should consider the double of the maximum of (1) the length of the trajectory joining the particle A to the particle B, passing through their common source, and (2) the length of the trajectory joining the particle C to the particle D, passing through their common source.

Last but not least, one should also have in mind that, according to special relativity, simultaneity is not absolute and is a frame dependent concept. This opens the door to other types of experiments that we will not discuss here however [20].

### 3.3 An a priori prediction ?

One may wonder whether there are any a priori theoretical estimates of a distance scale D or a propagation velocity V for non-standard correlations. propagation speed V of non-standard correlations. By playing only with the usual fundamental constants h, c and G can only give V = c or V = ∞ and D $10^{-33}$ cm= Planck length which is excluded. Playing with less fundamental constants such as the quark mass $M_{quark}$ allows us to multiply these values by any power N of the dimensionless constant $GM/hc \approx 10^{-39}$. For small values of N, such as -1 or +1, we obtain V= $10^{\pm 39}$ cm and D = $10^{-33}$ cm, which are excluded, or $10^{39}$ cm, which is unobservable by a lunar experiment on the Earth-Moon distance scale. Another source of a priori estimates for the distance scale D could come from the MOND theory [21] as an alternative theory to dark matter, which is characterised by the constant D ~10 kpc, not observable onthe Earth-Moon distance scale. As can be seen, the range of possible a priori predictions has no firm constraints, and only experimentation can potentially provide a constraint. None of these estimates leads to a distance up to about 10 times the Earth-Moon distance, which is theonly scale at which a deviation from the Earth-Moon distance can beobserved. The detection of a deviation from the prediction of quantum mechanics would therefore lead to a new fundamental constant.

## 4 Various experimental configurations

### 4.1 Estimating a lower bound on the "speed of collapse" in Cao's Earth-Satellite experiment

The present record on the distance over which Bell's inequalities have been violated is possessed by Cao's team who managed to measure quan- tum correlations, in a Quantum Key Distribution scheme à la Ekert, between two entangled photons emitted by a satellite and measured in two chinese cities separated by a distance of 1203 kilometer [14]. The satellite was placed on an orbit at 500 kilometer above Earth's surface. We may thus estimated the distance between the satellite and each of the cities to be of the order of square root of two times 500 kilometer, say 700 kilometer, which leads to bound on the speed of quantum communication of the order of times $10^7 \cdot c$ (more or less fifteen times the bound associated to Gisin's experiment [18].

### 4.2 Estimating a lower bound on the "speed of collapse". In Earth-Moon experiments.

Several experimental configurations are possible. A first one in which the source lies at the surface of Earth (figure 2 case 1) with one detector on the moon was proposed in 2009 [25, 26]. An other option is possible where one of the photons reflects on a mirror placed at the surface of the Moon (figure 2 case 2); an initial proposal was inspired by the common practice of "Earth-Moon lasers", in which a laser beam is sent to a mirror installed on the Moon and the reflected beam is detected on Earth. In this case, the aim was to create entangled pairs of photons using a laser and to measure the correlation of polarisation for photons reflected by the Moon and those remaining on Earth [24, 26].

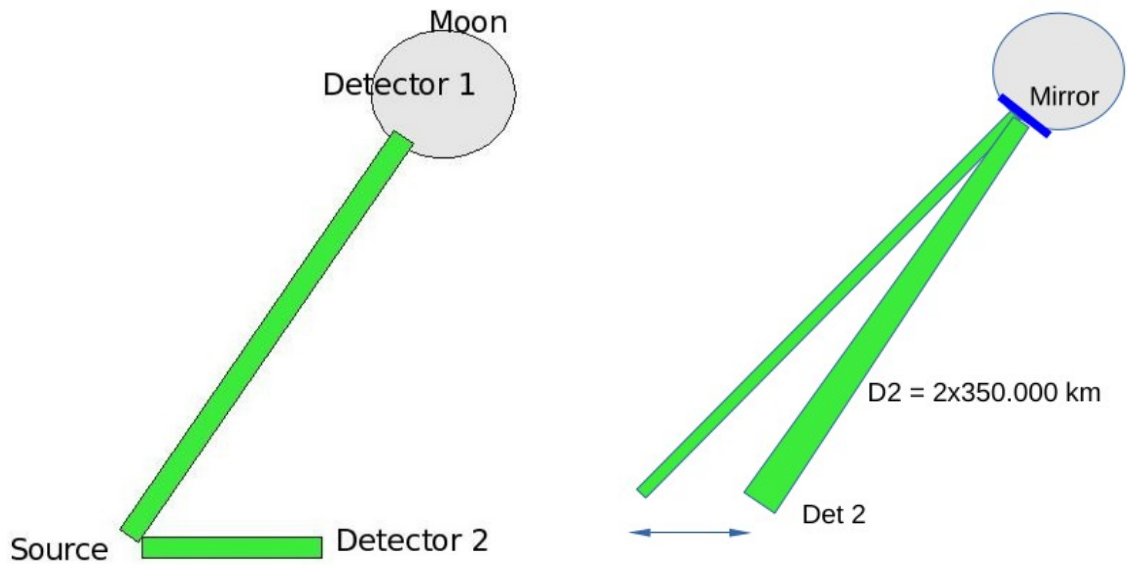

Figure 2 : Source on the Earth, case 1 ; source on the Earth : case 2

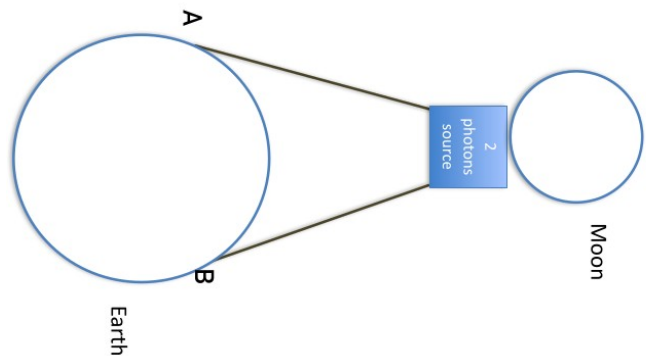

Figure 3 : Source on the Moon

The third case (figure 3) corresponds to a situation where the source is placed at the surface of the moon. They would lead to a relative gain of more or less 300, in estimating a lower bound on the speed of quantum correlations, relative to the chinese earth-satellite experiment [14].
Recently, a variant has been proposed [27] with spacecrafts located at Lagrange points (figure 4). In this approach one would install orientable polarizers in two spacecrafts and a source of correlated pairs of photons on the third spacecraft. Compared to Earth-Moon correlations the gain
in distance would be a factor 20.

The set ups described in figures 2 case 2 and 4 would present very appealing advantages regarding quantum key distribution, compared to the case where the source of entangled photons is put in an artificial satellite as in the chinese experiment [14]. This is so because all points located at the surface of Earth would be accessible, and the duration of the transmission is quite longer in this configuration.

As mentioned by Cao et al. [27], the configuration with one detector on Earth suffers from the timing perturbation and laser beam dispersion caused by the Earth atmosphere. We thus suggest a new proposal where we put the photon sources on the Moon, which can be more intense than a source on a satellite at L4/L5, and the photon detectors on L4 and L5. The perturbations due to the Earth atmosphere can then be avoided. This is similar to the LISA project of the European Space Agency where laser links will exist between three spacecrafts separated by 5 millions km. Our proposal is then to use polarized lasers and to install orientable polarizers in two spacrafts and a source of correlated pairs of photons on the third spacecraft. Compared to Earth-Moon correlations the gain in distance would be a factor 20. In principle one has also to take into account the general relativistic gravitational difference of the proper time $\alpha = 1 - GM/Rc^2$ for the Earth and the Moon, where G is the gravitational constant, M and R the mass and radius on the celestial object and c the speed of light. For the Earth and the Moon one has respectively $\alpha = 1 - 0.08$ and $\alpha = 1 - 0.0031$.

Therefore, for detection cadence large than $1/0.08 = 12$ photons/sec this correction has to be accounted for.

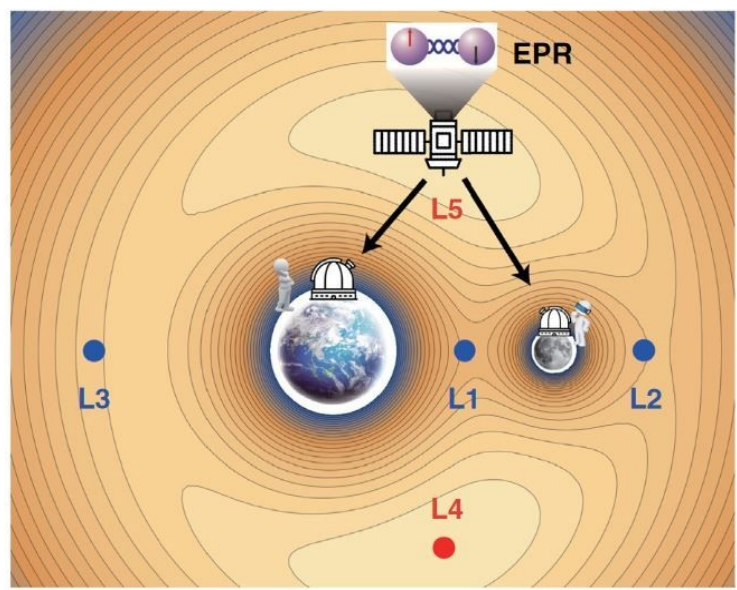

Figure 4: Source beyond the Earth-Moon system with one detector on the moon and another one on Earth. L represents Lagrange points (re-produced with the courtesy of the authors of ref. [27]).

### *4.3 Going beyond the moon*
Beyond the Moon, Kaltenbaek et al [28] have already proposed to make correlation experiments with a Martian base. The advantage is the gain in distance (a factor of the order of 1000 compared to Earth-Moon correlations), But a complex Martian base required for such an experiment cannot be foreseen before many decades.

## 5 Conclusions

As pointed out by Louis de Broglie repeatedly, since his seminal paper of 1927, but also later with Andrade e Silva [6] and actually for the rest of his life [8], de Broglie tried to get rid of the configuration space. He was the first to understand that the necessity to describe entangled states in configuration space is deeply counterintuitive, and is the signature of what Einstein called spooky action-at-a-distance, that we call nowadays non-locality. For instance, in the 50's, he wrote the following [2]: ..."*Or, la méthode de Schrödinger implique nécessairement l'emploi de l'espace de configuration et ne permet pas de se représenter le phénomène physique constitué par le mouvement des corpuscules dans le cadre de l'espace physique. Sans doute la Mécaniqueclassique se servait-elle souvent, elle aussi, de l'espace de configuration, mais ce n'était pas pour elle une nécessité: elle pouvait raisonner en considérant le mouvement des points matériels du système dans l'espace à trois dimensions et elle n'employait l'espace de configuration que comme un artifice mathématique permettant de présenter plus élégamment ou d'effectuer plus aisément certains calculs. Dès l'apparition des Mémoires de Schrödinger, tout en reconnaissant l'exactitude des résultats obtenus par sa méthode, j'avais trouvé paradoxal le principe même de cette méthode*"... In ref.[29] he added: ...*the fictitious space has never satisfied me, and I have done great work on this. In particular, one of my students, Mr Andrade e Silva, did a doctoral thesis to show how one can interpret this with our ideas, that is to say that everything happens in a physical space and it is only a certain representation in the configuration space...*

From this point of view, it is worth testing non-standard models of non-locality, for instance quantum aether models à la Vigier [10, 17], or models in which quantum information propagates at finite speed in 3 D space. These models share as a common feature that 3 D space constitutes a memory [30] of (the actualisation of) quantum correlations (in the same vein as Sheldrake's hypothesis[4] of shape waves which would impregnate space-time).

From this point of view, violating Bell's inequalities at interplanetary distances is a challenging and stimulating program.

This quest is stimulating from the point of view of foundations (will entanglement survive at these distances?) but also from a technological point of view: if the correlations do not fade away, their use paves the way to a quantum key distribution channel efficient at arbitrary locations on Earth (figure 3).

This program is hard to realize however if we note that, as mentioned in ref.[32], ...*Unlike the exponential attenuation in fiber links, the dominant loss in satellite-based QKD arises from geometric optics effects: beam divergence through the atmosphere coupled with finite receiving telescope aperture limitations. And the geometric optics effects (scale like) $L^{-2}$ where L is the length of the link....*

Many technical difficulties have already been solved in the past however, and there is growing interest for developing constellations of satellites providing secure QKD links, among others at the european level [33, 34]. We expect that new records will be broken in a near future, which is promising both at fundamental and applied levels.

---

[4] Rupert Sheldrake assumed [31] that the evolution of living organisms at the surface of the earth would be accelerated by a kind of cosmic memory that would gather and disseminate (non-locally) at the scale of the planet the teachings brought by experiences lived by individual organisms. That such a wave exists at the level of living organisms is still an open question. R. Sheldrake also emitted the hypothesis according to which the de Broglie frequency that is equal to the mass-energy of a quantum object divided by the Planck constant expresses the level of self-memory of this object. Inertia, or resistance to a change of position, that is also proportional to the mass-energy would reveal the tendency of an object to "repeat itself", or to stay equal to itself.

## 6 Acknowledgment.
One of the authors (T.D.) acknowledges the support of the EU: the EIC Pathfinder Challenges 2022 call through the Research Grant 101115149 (project ARTEMIS). One of us (J.S.) acknowledges authorization by the authors of ref. [27] to reproduce picture 4 from their paper.